\documentstyle[preprint,aps,prl,epsf]{revtex}
\newcommand{\be}{\begin{equation}}
\newcommand{\ee}{\end{equation}}
\newcommand{\ben}{\begin{eqnarray}}
\newcommand{\een}{\end{eqnarray}}

\begin{document}
\author{L. Diambra \thanks{Electronic address: diambra@fisio.icb.usp.br} }
\address{Departamento de Fisiologia e Biof\'{\i}sica,
Universidade de S\~ao Paulo \\ Av. Prof. Lineu Prestes 1524, ICB1
cep 05508-900, S\~ao Paulo, SP Brasil.}
\title{Divergence Measure Between Chaotic Attractors}
\maketitle
\begin{abstract}
We propose a measure of divergence of probability distributions
for quantifying the dissimilarity of two chaotic attractors. This
measure is defined in terms of a generalized entropy. We
illustrate our procedure by considering the effect of additive
noise in the well known H\'enon attractor. Comparison of two
H\'enon attractors for slighly different parameter values, has
shown that the divergence has complex scaling structure. Finally,
we show how our approach allows to detect non-stationary events
in a time series.
\end{abstract}
\newpage
Through the appropriate embedding procedures, strange attractors
can be numerically approximated by a large sets of points, either
from experimental time series (TS) or from numerical simulation
of chaotic systems. Advances in nonlinear analysis of TS have
made possible to identify and classify chaotic dynamical systems,
to determine if a signal is deterministic or not, and to
establish correlations where the traditional linear analysis were
not sensitive. However, there are many situations where we need
not a complete characterization of an attractor, but rather a
quantitative way of comparing attractors. For instance, recently
several authors have proposed to use some measures of
dissimilarity of attractors to analyze non-stationary signals
\cite{SM96,S97} and for TS classification \cite{SS97}. In some
situations it could be important to quantify the difference of
two attractors from the same chaotic dynamical systems,
corresponding to slightly different parameters. The computation
of the hierarchy of generalized dimensions does not help, because
even if all dimensions of two fractal sets are equal, this does
not guarantee that the two fractal objects are identical. In
order to give a quantitative answer to these issues, a number of
dissimilarity measures have been proposed in the literature
\cite{S97,hk94,arp}. The quantitative comparison of attractors
can be relevant in many different problems such as: numerical
taxonomy of TS; to establish a criterion for stationarity; to
study the numerical convergence of chaotic solutions; to evaluate
the effect of nonlinear noise reduction of noisy chaotic
attractors, among other applications. For the above mentioned
purposes we need a reliable way of comparing attractors rather
than their detailed characterization. In this paper, we propose a
divergence measure based on a generalized entropy function for
quantifying the similarity of attractors.

From the information theory viewpoint, the amount of uncertainty
of the probability distribution (PD), ${p_i} $, is defined in a
general way by $ H_f\left[ p_i \right] =-\sum _i f\left[ p_i
\right] $ \cite{daro75}. There is not a unique information measure
$H_f$. The more commonly used information measure or entropy
function was introduced by Shannon \cite{sh} where $f\left( p
\right) = p \ln p $. Generalized entropy $f_q $ has been
postulated by R\'enyi \cite{rey} and Havrda-Charv\'at \cite{hch}.
R\'enyi's generalized entropy has been used to define a hierarchy
of generalized dimensions \cite{GP}. Tsallis introduced the
Havrda-Charv\'at entropy function to elaborate an non-extensive
thermodynamics \cite{tsa}. Associated with an entropy function
$f$, we have a divergence measure $D_f\left( p : \hat{p} \right)$
between two PD ${p_i} $ and $\hat{p}_i $. A general divergence
measure form associated to the $f$-entropy was given by Csisz\'ar
\cite{csis}
\begin{equation}
D_f\left( p: \hat{p}\right) =\sum_i \left( \hat{p}_i f\left[
\frac{p_i}{\hat{p}_i} \right] + p_i f\left[ \frac{\hat{p}_i}{p_i
}\right] \right) , \label{distance}
\end{equation}
where $f $ is a convex function and one imposes the condition
$f\left( 1\right) =0$, that guarantees $D_f\left( p : p \right)
=0$. R\'enyi's generalized entropy does not satisfy convexity.
Fortunately, Havrda-Charv\'at entropy function fulfills this
property. For this reason we will work here with the
Havrda-Charv\'at entropy function. From here on we shall refer to
the generalized divergence measure associated with the
Havrda-Charv\'at entropy function, simply as the $q$-divergence,
and will be denoted by $D_q$. If we replace $f$ by the function
corresponding to the Shannon entropy, we obtain the well known
Kullback-Leibler distance \cite{k1}.
\begin{equation} D_{1}\left( p:\hat{p}\right) =\sum_i
\left( p_i \ln \left[ \frac{p_i }{\hat{p}_i }\right] + \hat{p}_i
\ln \left[ \frac{\hat{p}_i }{p_i }\right] \right).
\label{kullback}
\end{equation}
The function $f$ corresponding to the Havrda-Charv\'at entropy is
given by $f_q \left( p \right) = \left( q -1\right) ^{-1}\left(
p^q -p \right) $. Then, the associated $q$-distance is given by
\begin{equation}
D_q\left( p:\hat{p}\right) =\left( q -1\right) ^{-1} \left( \sum
_i p_i^q \hat{p}_i ^{1-q} + \sum _i \hat{p}_i^q p_i^{1-q} -2
\right) . \label{div}
\end{equation}
It is easy to show that $D_q\left( p :\hat{p} \right) \rightarrow
D_{1}\left( p :\hat{p} \right) $ when $q \rightarrow 1$. The
$q$-divergence measure is positive definite, has been made
symmetric, and fulfills $D_q\left( p : p \right) =0$. Also
$D_q\left( p :\hat{p} \right) $ considered as a function of
${p_i} $ and ${\hat{p_i}} $ is convex. We remark that $D_q$ is
semi-metric, since it may not satisfy the triangular inequality.

After these definitions, let us consider the divergence between
two finite time series embedded in ${\cal R}^d$, $X=\left( {\bf
x}_0, \ldots ,{\bf x}_N \right) $ and $Y=\left( {\bf y}_0, \ldots
,{\bf y}_N \right) $. There are two well known ways of estimating
the quantities (\ref{kullback}) and (\ref{div}) from $X$ and $Y$.
The most straightforward, but also more expensive, is to use a box
counting approach: one defines a partition $\Pi _{\varepsilon}$ of
the state space, with characteristic size $\varepsilon $. Thus the
probability to find a point of $X$ ($Y$) in the $i$th box is
$p_i$ ($\hat{p} _i$). By counting the number of points $n_i$ in
the box $i$, the probabilities $p_i$ can be estimated as
$p_i=n_i/N$, where $N$ is the total number of points. Some
authors have refined this procedure, by adapting the size of the
boxes depending on the local density \cite{fra}.

On the other hand, a more efficient method of estimating
(\ref{kullback}) and (\ref{div}) is by correlation sums
\cite{gp}. Instead of taking a fixed mesh, one can calculate the
probability $P\left( {\bf x}_j, \varepsilon \right) $ to find a
$d$-dimensional point within a sphere of radius $\varepsilon $
centered around the ${\bf x}_j$, with $j=1,\ldots ,M$ randomly
chosen from the trajectory $X$. $P\left( {\bf x}_j ,\varepsilon
\right) $ is estimated by counting the number $n_j$ of points
falling in the sphere of radius $\varepsilon$ centered at ${\bf
x}_j$,
\begin{eqnarray}
P\left( {\bf x}_j, \varepsilon \right) =N^{-1} \sum _{i=1}^N
\Theta \left( \varepsilon -|{\bf x}_i-{\bf x}_j| \right) \ \ \ \
j=1,\ldots ,M. \nonumber
\end{eqnarray}
In our case the expressions (\ref{kullback}) and (\ref{div}) can
be rewritten as
\begin{equation} D_{1}\left( p:\hat{p}, \varepsilon \right) = M^{-1}
\left( \sum _{j=1}^M \ln \left[ \frac{\sum _{i=1}^{N^{\prime }}
\Theta \left( \varepsilon -|{\bf y}_i-{\bf x}_j| \right)}{\sum
_{i=1}^{N^{\prime }} \Theta \left( \varepsilon -|{\bf x}_i-{\bf
x}_j| \right)} \right] + \sum _{j=1}^M \ln \left[ \frac{\sum
_{i=1}^{N^{\prime }} \Theta \left( \varepsilon -|{\bf x}_i-{\bf
y}_j| \right)}{\sum _{i=1}^{N^{\prime }} \Theta \left(
\varepsilon -|{\bf y}_i-{\bf y}_j| \right)}\right] \right),
\label{kul2}
\end{equation}
\begin{equation} D_q\left( p:\hat{p}, \varepsilon \right)
= \frac{M^{-1}}{\left( q-1\right) } \left( \sum _{j=1}^M \left(
\frac{\sum _{i=1}^{N^{\prime }} \Theta \left( \varepsilon -|{\bf
y}_i-{\bf x}_j| \right)}{\sum _{i=1}^{N^{\prime }} \Theta \left(
\varepsilon -|{\bf x}_i-{\bf x}_j| \right)}\right)^q  +  \sum
_{j=1}^M \left( \frac{\sum _{i=1}^{N^{\prime }} \Theta \left(
\varepsilon -|{\bf x}_i-{\bf y}_j| \right)}{\sum
_{i=1}^{N^{\prime }} \Theta \left( \varepsilon -|{\bf y}_i-{\bf
y}_j| \right)}\right)^q - 2 \right). \label{div2}
\end{equation}
Where $\Theta (x)$ is the step function which has the value 1 if
$x \geq 0$ and is 0 otherwise, $|{\bf x}_i-{\bf y}_j |$ is the
distance between ${\bf x}_i$ and ${\bf y}_j $. The sum is taken
only for those $i$'s and $j$'s that are separated in time by more
than $B$ sampling times to avoid artifactural correlations
\cite{th86}, thus $N^{\prime }=N-d-B$. Notice that the quantities
(\ref{kullback}) and (\ref{div}) are defined for two finite
discrete probability distributions $p_i$ and $\hat{p} _i$ only if
$p_i >0$, and $\hat{p} _i >0$, and if there is a one-to-one
correspondence between the elements $i$ \cite{rey}. In order to
satisfy these requirements, we perform the summation in the box
counting approach (Eq. (\ref{kullback}) and (\ref{div})) only over
the boxes $i$ that contain points from both $X$ and $Y$ (i.e. $p_i
>0$ and $\hat{p}_i >0$). In the sphere counting scheme (Eq.
(\ref{kul2}) and (\ref{div2})), we include in the summation over
$j$ only spheres which contain points both in $X$ and $Y$ and for
this reason $M$ decreases with $\varepsilon $.

Now, we shall present some examples of numerical computations of
the q-divergence $D_q$ between two time series. We estimate the
$D_q$ by means of the two before mentioned methods: the box
counting (BC), and the sphere counting (SC). In the case of BC
algorithm, for simplicity, we restrict our analysis to dimension
$d=2$. As our first example, we deal with trajectories of $10 \
000$ points of the H\'enon model $x_{n+1}= 1 - a x_n^2 +
bx_{n-1}$, with parameters $a=1.4$ and $b=0.3$; cf. Kantz
\cite{hk94}. The set $X$ corresponds to the clean attractor,
while the set $Y$ consists of the same set of points plus an
additive Gaussian noise. In Fig. 1 we present the mean value of
$D_1$ {\it versus} $\varepsilon $ computed with the BC method. We
computed the mean value over 5 realizations of $Y$ with
signal-to-noise ratio $\eta = 20$ dB \cite{SNR}. Of course, the
divergence depends of the length scale $\varepsilon $. By
choosing a relatively small $\varepsilon $, $D_q$ will pick up
local differences between $X$ and $Y$. However, taking
$\varepsilon $ too small leads to poor statistics. For large
$\varepsilon $, we lose the small scale structure of the
attractors that they become indistinguishable. We can see that
$D_1$ reaches a maximum at a values of $\varepsilon $ that will be
denoted by $\varepsilon _0$. The dashed curve corresponds to $S =
\langle D_1 \rangle /\sigma \left( D_1\right) $ where $\langle
\rangle $ denotes the mean value over 5 realizations and $\sigma$
denote the standard deviation. We can see that the maximum value
of $D_1$ presents very good statistics.

In Fig. 2 we display the $q$ dependence of $D_q \left(
\varepsilon _0 \right)$ for different noise level ($\eta =30$ dB
in solid line, $\eta =20$ dB in dashed line and $\eta =10$ dB in
dotted line). As shown in Fig. 2, the parameter $q$ is like a gain
control parameter. The divergence of the two attractors increases
with $q$. This fact can be used to detect small divergence as we
will discuss in the last example. In Fig. 3 we illustrate the
behavior of the mean value of $D_2$ {\it versus} $\varepsilon $
computed with SC method. We computed the mean value over 5
realization of $Y$ for each level of noise. We can see that $D_2$
presents a maximum at $\varepsilon _0$ which depends on the
characteristic length scale at which the attractors differ. The
$\varepsilon $ dependence of $D_q$ characterizes the relationship
between the sets $X$ and $Y$. In the case analyzed here, this
length scale is related to the level of noise added to $X$. In
Fig. 4 we display $D_2\left( \varepsilon _0 \right) $ {\it
versus} $\eta $. This figure shows that the $D_2\left( \varepsilon
_0 \right) $ scales exponentially with the level of noise added to
the signal.

Now, we compare two clean H\'enon attractors for slightly
different parameter values. In this case $Y$ corresponds to
parameters $a=1.35$ and $b=0.31$, while $X$ remains the same. In
Fig. 5 we display $D_1$ computed with $N=50 \ 000$ points,
embedding dimension $d=2$, using the BC method (dotted line),
using the SC method (dashed line), and with $d=3$ using the BC
method (solid line). When we compare these attractors, we find
that the divergence measure as a function of $\varepsilon $
exhibits a more complex scaling structure than showed in the
previous example. This complex relationship between the attractors
has not been reported in early studies \cite{hk94,DZTD96}.

Let us finally show how the $q$-distance $D_q$ can be used to
detect non stationarity events in a TS. As numerical example, let
us consider a generalized baker map defined by
\begin{eqnarray}
v_n \leq \alpha : \ \ \ \  u_{n+1} &=& \beta u_n, \ \ \ \ \ \ \ \
\ \ \ \ \ v_{n+1}=v_n/\alpha, \nonumber \\
v_n > \alpha : \ \ \ \  u_{n+1} &=& 0.5 + \beta u_n, \ \ \ \ \
v_{n+1}=\frac{v_n-\alpha }{1-\alpha }. \nonumber
\end{eqnarray}
For this map, the parameter $\beta $ can be varied without
changing the positive Lyapunov exponent. We generate a non
stationary TS of length 8192 points with $\alpha =0.4 $ and two
values of $\beta $. In the first 4096 iterations we use $\beta
=0.6$, and in the second 4096 iterations we set $\beta =0.8$. We
record $u+v$, then we subtract the mean value and normalize to
unit variance separately each one of the two parts; cf. Schreiber
\cite{S97}. Thus, we have a non stationary event in the middle of
the TS that is very hard to detect, because observables like
mean, variance and maximal Lyapunov exponent are constant by
construction. Figure 6 shows $D_2\left(S_i:S_j\right) $ and
$D_6\left(S_i:S_j\right) $, between two nearest non-overlapping
segments $S_i$ and $S_j$ of 1000 points \cite{note}. This example
shows also that the parameter $q$ plays a role of a nonlinear gain
parameter. We can see that for $q=6$ the divergence was able to
detect precisely when the small change in $\beta $ ocurr, while
for $q=2$ we have a poor discrimination power. We used resolution
$\varepsilon =0.1155$ in the computation. We have a window
resolution with similar results for $\varepsilon $ in $[
0.075,0.15] $.

We have introduced the $q$-divergence as a measure of
dissimilarity of two finite sets. Our approach is particularly
useful for comparing attractors. We found that the divergence
decreases exponentially when $\eta $ increases. Comparison of two
H\'enon attractors for slighly different parameter values has
shown that the $q$-divergence has a complex scaling structure.
Also this tool promises to be useful for detecting non-stationary
event in a TS, even in very hard conditions. Thus It will be seen
that some interesting physical insight is gained by recourse to
this type of dissimilarity measure.

The author acknowledges the financial support of FAPESP Grant
N$^o$ 99/07186-3. The author thank C.P. Malta for a careful
reading of the manuscript and her useful comments.

\begin{figure}[h]
\epsfxsize=9.5cm \epsfysize=7cm \epsffile{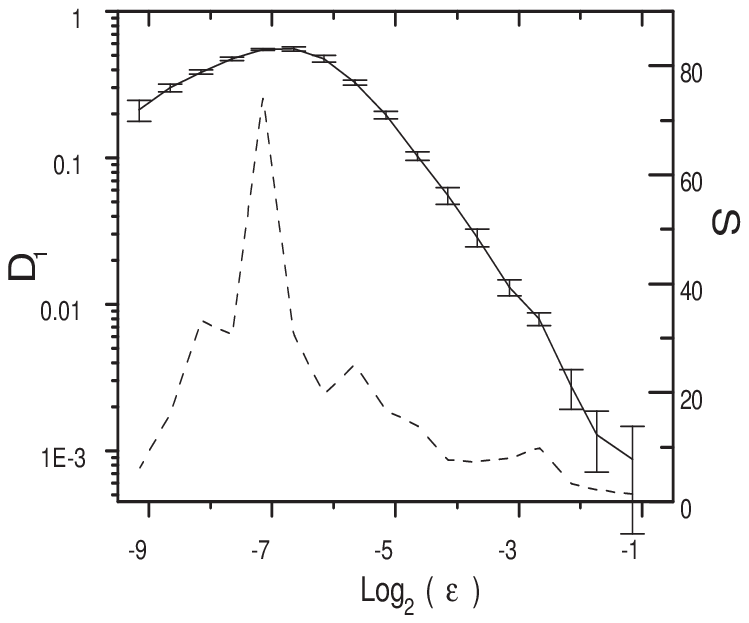}
\caption{Solid Line: $D_1\left( \varepsilon \right)$ between the
TS $X$ generated by the H\'enon system and the TS $Y$ generated by
the same systems contaminated with noise ($\eta = 20$ dB),
computed with the BC scheme using $N=10 \ 000$ points and $d=2$
(left axis). Dashed line: the statistic $S \left( \varepsilon
\right)$ (right axis).}
\end{figure}

\begin{figure}[h]
\epsfxsize=9cm \epsfysize=7cm \epsffile{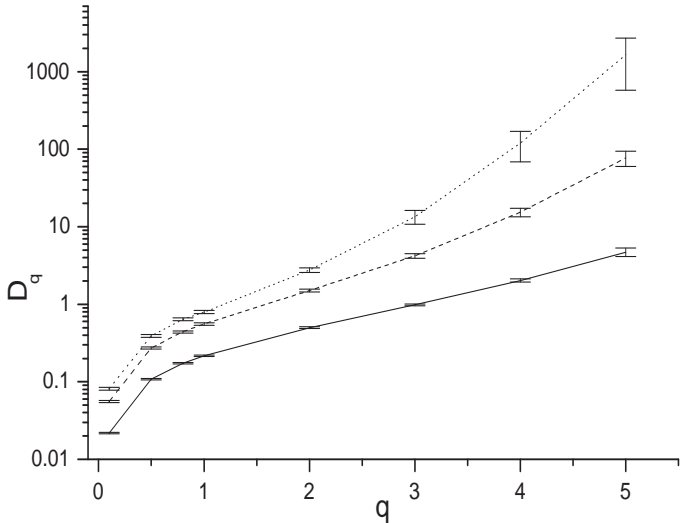} \caption{The
$q$-distance $D_q\left( \varepsilon _0 \right)$ between the TS
$X$ generated by the H\'enon system and the TS $Y$ generated by
the same systems contaminated with several level of noise as a
function of the parameter $q$ (solid line: SNR= 30 dB, dashed
line: $\eta = 20$ dB and dotted line $\eta = 10$ dB) . The
calculation were performed with the BC scheme using $N=10 \ 000$
points and $d=2$.}
\end{figure}
\newpage
\begin{figure}[h]
\epsfxsize=9cm \epsfysize=7cm\epsffile{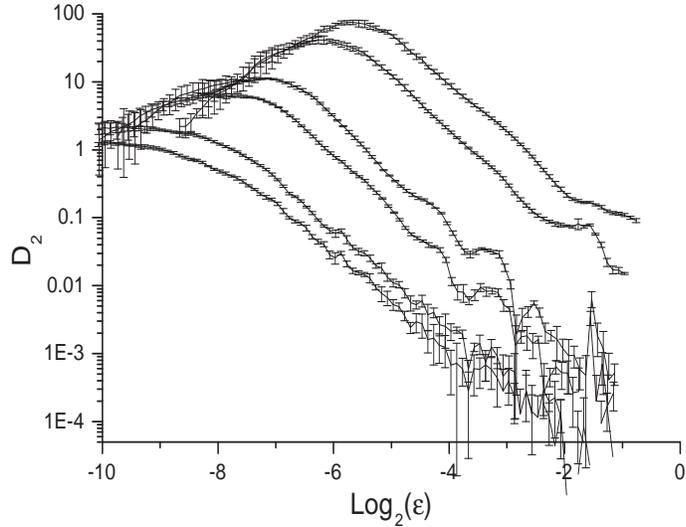} \caption{
Distance $D_2$ between H\'enon system and itself contaminated
with several level of noise (between $\eta = 30$ dB to $\eta = 7$
dB). The calculation were performed with the SC scheme using $N=10
\ 000$ points, $B=20$ and $d=2$.}
\end{figure}
\begin{figure}[h]
\epsfxsize=9cm \epsfysize=7cm \epsffile{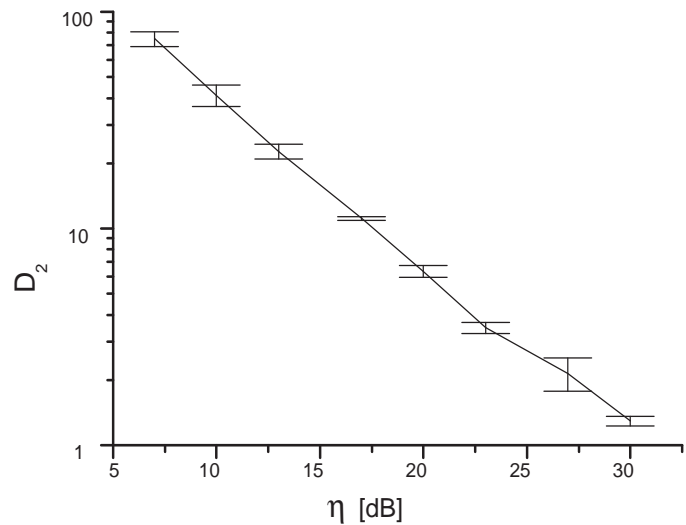}
\caption{Divergence $D_2\left( \varepsilon _0 \right)$ between
H\'enon system and itself contaminated with noise as a function
of the level of noise $\eta $.}
\end{figure}
\newpage
\begin{figure}[h]
\epsfxsize=9cm \epsfysize=7cm \epsffile{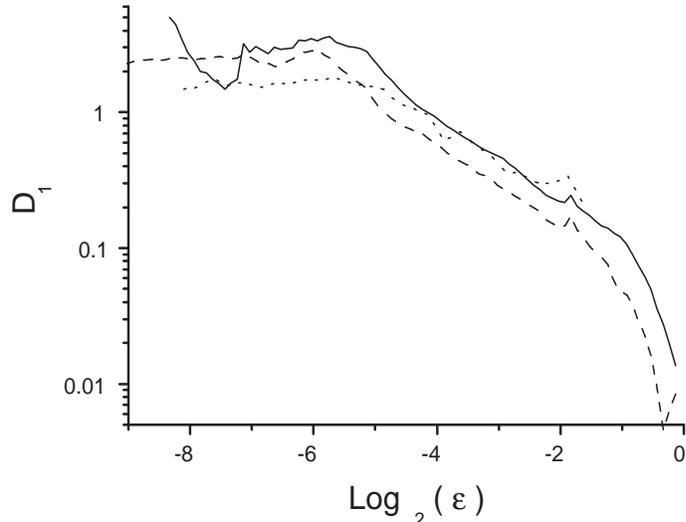}
\caption{Estimated values of $D_1$ as a function of $\varepsilon
$ for two time series of length $N=50 \ 000$, and $B=20$
generated by slightly different H\'enon systems. Dotted line:
using BC method with $d=2$. Dashed line: using SC method with
$d=2$. Solid line: using SC method with $d=3$.}
\end{figure}

\begin{figure}[h]
\epsfxsize=9cm \epsfysize=7cm \epsffile{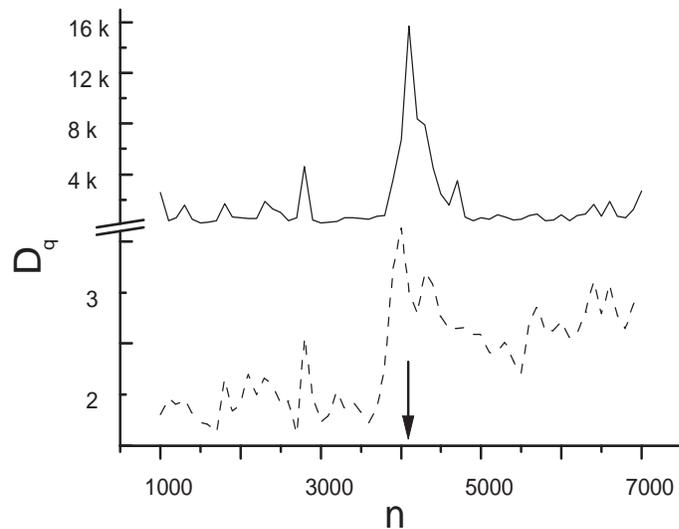} \caption{$D_6$
(solid line) and $D_2$ (dashed line) between two non-overlapping
subsequent segments of TS with a non-stationary event at 4096.}
\end{figure}


\begin{references}
\bibitem{SNR} The signal-to-noise ratio $\eta $ in decibels (dB) is
defined as $10 \log _{10} \left[ \sigma ({\rm signal})/\sigma (
{\rm noise}) \right]$.
\bibitem{note} The total signal of 8192 points is split into 71
segments of 1000 points with 900 points of overlapping. Thus
$j=i+10$.
\end{references}
\end{document}